\documentclass[aps,prd,showpacs,letterpaper,superscriptaddress,nofootinbib,twocolumn]{revtex4}
\usepackage{bm}

\usepackage[latin1]{inputenc}
\usepackage[english]{babel}
\usepackage[dvips]{graphicx}
\usepackage{amssymb}
\usepackage{epsfig}

\newcommand{\be}{\begin{equation}}
\newcommand{\ee}{\end{equation}}
\newcommand{\bea}{\begin{eqnarray}}
\newcommand{\eea}{\end{eqnarray}}

\newcommand{\nd}{\noindent}
\newcommand{\la}{\lambda}
\def\lsim{\mathrel{\rlap{\lower4pt\hbox{\hskip1pt$\sim$}}\raise1pt\hbox{$<$}}}
\def\gsim{\mathrel{\rlap{\lower4pt\hbox{\hskip1pt$\sim$}}\raise1pt\hbox{$>$}}}
\def\nostrocostruttino#1\over#2{\mathrel{\mathop{\kern 0pt \rlap
{\hbox{$#1$}}} \hbox{\kern-.135em $#2$}}}

\begin{document}
%%%%%%%%%%%%%%%%%%%%%%%%%%%%%%%%%%%%%%%%%%%%%%%%%%%%%%%%%%%%%%%%%%%%%%%%%%%%%%

\title{Single spin asymmetries in inclusive hadron production \\
from SIDIS to hadronic collisions: universality and phenomenology }

\author{M.~Boglione}
 %\email{boglione@to.infn.it}
 \affiliation{Istituto Nazionale di Fisica Nucleare, Sezione di
 Torino, Via P. Giuria 1, I-10125 Torino, Italy}

\author{U.~D'Alesio}
 %\email{umberto.dalesio@ca.infn.it}
 \affiliation{Dipartimento di Fisica, Universit\`a di Cagliari, Cittadella Universitaria,
 I-09042 Monserrato (CA), Italy}
 \affiliation{Istituto Nazionale di Fisica Nucleare, Sezione di Cagliari, C.P. 170,
 I-09042 Monserrato (CA), Italy}

\author{F.~Murgia}
 %\email{francesco.murgia@ca.infn.it}
 \affiliation{Istituto Nazionale di Fisica Nucleare, Sezione di Cagliari, C.P. 170,
 I-09042 Monserrato (CA), Italy}

%\vspace{0.5cm}

\date{\today}

\begin{abstract}
\noindent In a perturbative QCD approach, with inclusion of spin and transverse momentum
effects, experimental data on azimuthal asymmetries observed in polarized semi-inclusive
deeply inelastic scattering and $e^+ e^-$ annihilations can be used to determine the Sivers,
transversity and Collins soft functions. By using \emph{these} functions, within the same
scheme, we predict $p^\uparrow p\to h+X$ single spin asymmetries in remarkable agreement with
RHIC experimental data.
\end{abstract}


\maketitle

Several azimuthal spin asymmetries have recently been measured in
different processes: semi-inclusive deeply inelastic scattering
(SIDIS)~\cite{Airapetian:2004tw,Alexakhin:2005iw,Ageev:2006da},
inclusive particle production from hadronic
collisions~\cite{Adams:1991cs, Adams:1994yu, Bravar:1996ki,
  Adams:2003fx, Adler:2005in, Lee:2007zzh,Nogach:2006gm}
and from $e^+e^-$ annihilations~\cite{Abe:2005zx}. Sizable
transverse single spin asymmetries (SSAs) and large transverse
hyperon polarizations, see e.g.~\cite{Heller:1996pg}, severely
challenged the predictions of leading order perturbative QCD (pQCD)
obtained in collinear configuration~\cite{Kane:1978nd}, where the
constituent partons and the corresponding observed hadrons are
assumed to be collinear. Lately, a remarkable success in reproducing
SSA experimental data has been achieved in the framework of
non-collinear pQCD, by including spin and partonic intrinsic motion
effects. This requires the introduction of a new class of transverse
momentum dependent (TMD) parton distribution (PDFs) and
fragmentation functions (FFs). Phenomenologically, the most relevant
are, at leading twist, the Sivers~\cite{Sivers:1989cc,Sivers:1990fh}
and Boer-Mulders~\cite{Boer:1997nt} distribution functions and the
Collins fragmentation function~\cite{Collins:1992kk}. In fact, in a
factorized pQCD approach, they are responsible for most of the
azimuthal spin asymmetries observed in SIDIS and Drell-Yan (DY)
processes, and in $e^+e^-$ annihilations. Factorization theorems
have been proved for SIDIS and DY processes~\cite{Ji:2004xq}, in the
kinematical regime where two well distinct energy scales are
present: a large scale - either the photon virtuality or the
invariant mass of the leptonic pair, $Q$ - and a low-moderate scale
- the transverse momentum, $p_T$, of the observed hadron or of the
leptonic pair with respect to the colliding beams - so that
$\Lambda_{\rm QCD}\simeq p_T\ll Q$.

Although the large SSAs observed in inclusive single particle
production in hadronic collisions motivated a huge theoretical
effort towards a pQCD treatment of these processes with the
inclusion of TMD effects, a proof of factorization for this case is
still lacking. Moreover, it has been shown that in a QCD approach
the Sivers and Boer-Mulders effects vanish unless initial and final
state interactions between the struck parton and the spectators are
included~\cite{Brodsky:2002cx,Collins:2002kn}. These effects, which
can be accounted for by including appropriate color gauge links
(Wilson lines) in the invariant definition of the factorized
hadronic correlators, can actually be reabsorbed in prefactors
associated to the hard partonic cross sections. Notice that these
prefactors are different for different
processes~\cite{Bomhof:2006dp}. These results, while validating the
interpretation of the Sivers and Boer-Mulders functions as the basic
mechanisms generating SSAs and azimuthal asymmetries, question their
{\it universality} properties and the whole concept of
factorization. In other words, for these processes factorization
{\it is} broken, but in a known and predictable way (i.e.~through
the action of the hard scattering prefactors), so that the total
polarized and unpolarized cross sections can still be calculated in
a scheme where factorization should be understood at best in a
generalized, transverse momentum dependent way. Significant progress
in this sense has been achieved in the context of inclusive almost
back-to-back two-particle (e.g., dijet, photon-jet) production in
hadronic collisions~\cite{Vogelsang:2007jk, Collins:2007jp,
Bomhof:2007xt}. While these results give very useful indications and
suggest caution in treating single particle production at large
$p_T$, they cannot give definitive statements on this process, where
the low energy scale (the intrinsic $k_\perp$) is integrated over
and unobserved.

Alternatively, a more phenomenological approach to inclusive single
particle production in hadronic collisions has been formulated in
the context of a generalized parton model, with the inclusion of
spin and transverse momentum effects~\cite{Anselmino:1994tv,
Anselmino:2005sh}. In this approach factorization is assumed as a
reasonable starting point, leading twist TMD distributions keep
their partonic interpretation and are therefore expected to be
universal and process independent. Although reasonable, these
assumptions must undergo phenomenological scrutiny by careful
comparison with experimental measurements. In this scheme, one
should try to reproduce the available data on SSAs coming from
SIDIS, hadronic collisions and $e^+e^-$ annihilations with a {\it
single set} of universal, TMD distributions. Historically, this
approach was first applied to the $p^\uparrow p\to\pi+X$ process,
since low energy results were available at $\sqrt{s}=20$ GeV from
the E704 Collaboration. Later on, high energy regimes became
available at RHIC $\sqrt{s}=200$ GeV experiments. Recently, the same
approach has been independently applied to the Sivers and Collins
azimuthal asymmetries measured by the HERMES and COMPASS
Collaborations, and to the double hadron azimuthal correlations
measured in $e^+e^-$ annihilations. High quality global fits to all
experimental data have allowed the simultaneous determination of the
Sivers~\cite{Anselmino:2005ea} and transversity distributions and of
the Collins fragmentation function~\cite{Anselmino:2007fs}.

In this paper, for the first time, we investigate the possibility of reproducing unpolarized
cross sections and SSAs in proton-proton collisions, by using the Sivers and transversity
distribution functions, and the Collins fragmentation function, as determined by fitting
SIDIS and $e^+e^-$ experimental data, without any modification. No fit to $pp$ collision data
will be used nor performed, so that all our results may be considered as genuine predictions.

\begin{center}
\begin{figure*}
\hspace*{-20pt}
\epsfig{figure=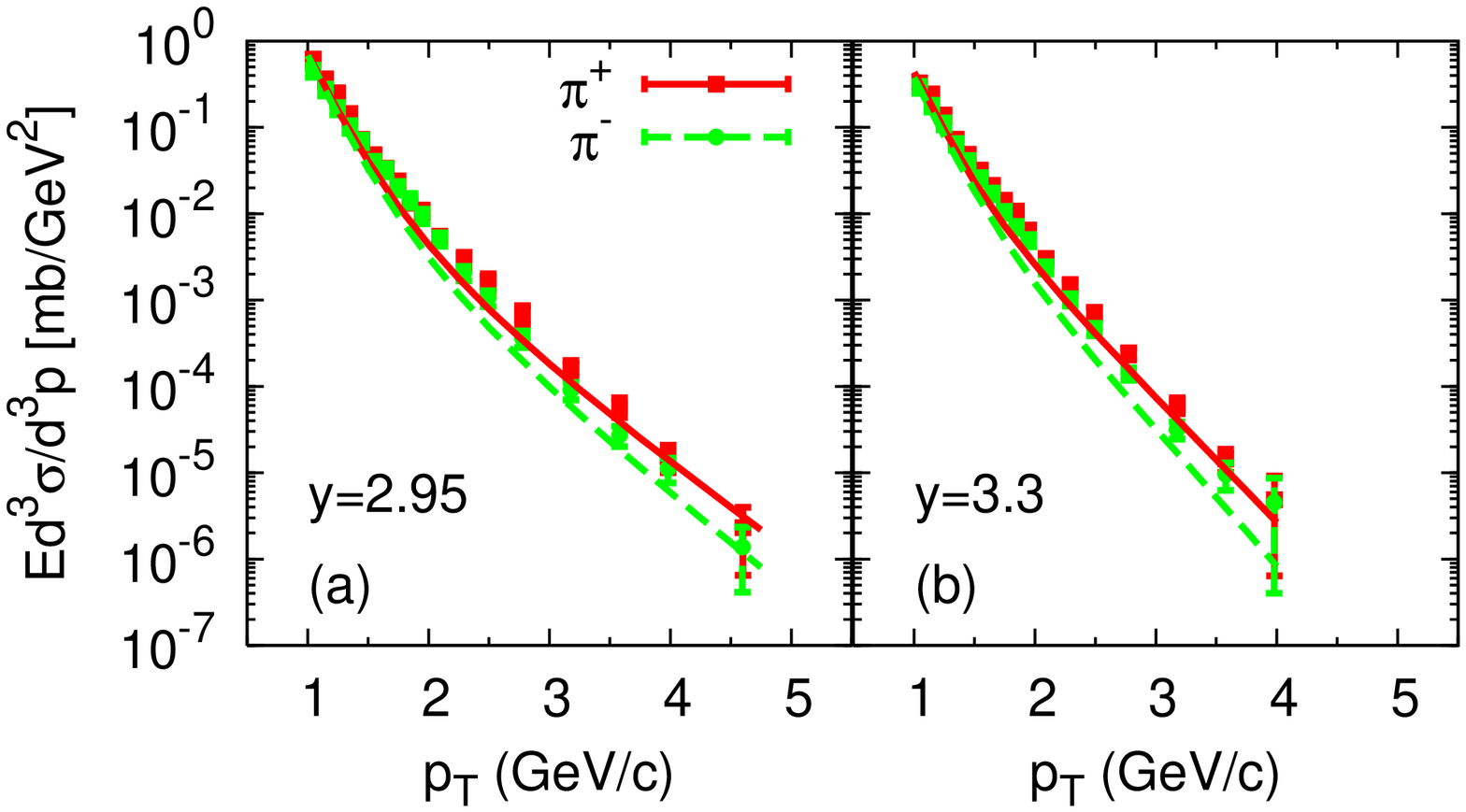,scale=0.43, angle=-90,bb=
540 50 560 640} \hspace*{10pt}
\epsfig{figure=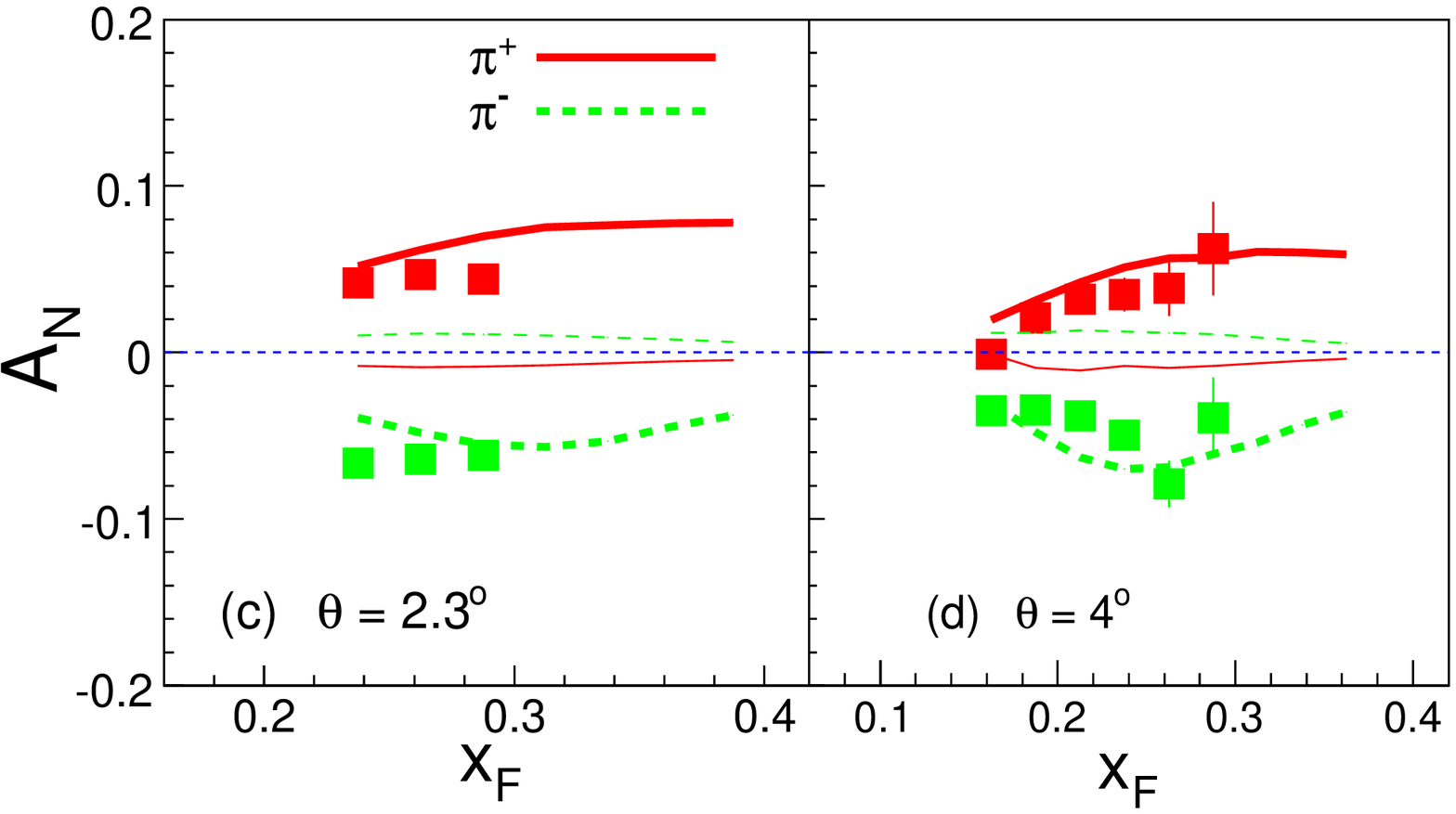,scale=0.38}
\caption{\label{brahms-pi} Panels (a) and (b): Invariant
differential cross section for $pp \to \pi^\pm +X$ at $\sqrt{s}=
200$ GeV  and two pseudorapidity values, $y=2.95$ and $y=3.3$,
vs.~$p_T$. Data are from BRAHMS~\cite{Arsene:2007jd}. Curves are
obtained adopting the same choices as in Fig.~1(a). Panels (c) and
(d): $A_N$ for $pp \to \pi^\pm  +X$ at $\sqrt{s}= 200$ GeV and two
different scattering angles, $\theta=2.3^\circ$ and
$\theta=4^\circ$, vs.~$x_F$. Data are from
BRAHMS~\cite{Lee:2007zzh}. Thick curves are obtained adding the
Sivers effect, as extracted from SIDIS data in
Ref.~\cite{Anselmino:2005ea}, and the Collins effect coupled with
the transversity function, as extracted from a global fit of SIDIS
and $e^+e^-$ data in
   Ref.~\cite{Anselmino:2007fs}. The thin lines show the Collins
   effect: notice that its sign is opposite w.r.t.~the Sivers contribution.}
\end{figure*}
\end{center}

Let us briefly sketch the basic relations adopted in our approach. For a detailed and
complete treatment we refer to the original papers~\cite{D'Alesio:2004up, Anselmino:2005sh}.
The invariant differential cross section for the polarized process $p^{\uparrow,\downarrow}
p\to\pi+X$ can be written as
\begin{widetext}
 \bea
 \frac{E_\pi \, d\sigma^{p^{\uparrow,\downarrow} p \to \pi + X}}
{d^{3} \bm{p}_\pi}
 &= & \sum_{a,b,c,d, \{\la\}}
 \! \int \frac{dx_a \,
dx_b \, dz}{16 \pi^2 x_a x_b z^2  s} \; d^2 \bm{k}_{\perp a} \, d^2 \bm{k}_{\perp b}\, d^3
\bm{k}_{\perp\pi}\, \delta(\bm{k}_{\perp\pi} \cdot \hat{\bm{p}}_c) \, J(\bm{k}_{\perp\pi})
\label{gen} \\
&\times&  \rho_{\la^{\,}_a, \la^{\prime}_a}^{a/p^{\uparrow,\downarrow}} \, \hat
f_{a/p^{\uparrow,\downarrow}}(x_a,\bm{k}_{\perp a}) \> \rho_{\la^{\,}_b,
\la^{\prime}_b}^{b/p} \, \hat f_{b/p}(x_b,\bm{k}_{\perp b}) \, \hat M_{\la^{\,}_c,
\la^{\,}_d; \la^{\,}_a, \la^{\,}_b} \, \hat M^*_{\la^{\prime}_c, \la^{\,}_d; \la^{\prime}_a,
\la^{\prime}_b} \> \delta(\hat s + \hat t + \hat u) \> \hat
D^{\pi/c}_{\la^{\,}_c,\la^{\prime}_c}(z,\bm{k}_{\perp\pi})\,.\nonumber
 \eea
%\end{widetext}
%
The SSA can be expressed as the ratio $A_N=
(d\sigma^\uparrow-d\sigma^\downarrow)/(d\sigma^\uparrow+d\sigma^\downarrow)$. In
Ref.~\cite{Anselmino:2005sh} it was shown that in the TMD generalized parton model several
terms, other than those present in the collinear configuration, contribute to both the
numerator and
the denominator of $A_N$. %Eq.~(\ref{gen}).
These contributions were neglected in earlier TMD parton models which lacked proper inclusion
of spin effects. However, Refs.~\cite{D'Alesio:2004up, Anselmino:2005sh} have shown that for
both the E704 and RHIC kinematics, the numerator of the SSA is dominated by the Sivers
effect, with a small or negligible contribution from the Collins effect, while the
denominator is essentially given by the TMD unpolarized contribution which generalizes the
usual collinear one. Therefore, in the case of interest here, Eq.~(\ref{gen}) takes the
simple form
%
%\begin{widetext}
\bea
 \frac{E_\pi \, d\sigma^{p^{\uparrow,\downarrow} p \to \pi + X}}
{d^{3} \bm{p}_\pi}  & \simeq &
\sum_{a,b,c,d, \{\la\}}
 \! \int \frac{dx_a \, dx_b \, dz}{16 \pi^2 x_a x_b z^2  s} \; d^2 \bm{k}_{\perp a} \, d^2
\bm{k}_{\perp b}\, d^3 \bm{k}_{\perp\pi}\, \delta(\bm{k}_{\perp\pi} \cdot \hat{\bm{p}}_c) \,
J(\bm{k}_{\perp\pi})\\
&\times& \left [f_{a/p}(x_a,k_{\perp a}) \pm \frac{1}{2} \Delta^{\!N}\hat
 f_{a/p^{\uparrow}}(x_a,\bm{k}_{\perp a})\right ] f_{b/p}(x_b,k_{\perp b}) |\hat M_{\la_c,\la_d;\la_a, \la_b}|^2
 \> \delta(\hat s + \hat t
+ \hat u) \> D_{\pi/c}(z,k_{\perp\pi})\,. \nonumber
 \label{ansiv}
 \eea
\end{widetext}
Notice that the  contribution due to the convolution of the transversity
distribution with the Collins fragmentation function~\cite{Anselmino:2005sh}, has actually
been included in our calculations. However, this effect is marginal in our results, with the
exception of the BRAHMS $A_N$ of Fig.~\ref{brahms-pi}. All other terms are totally negligible
in the kinematical configurations considered, as their azimuthal phase factors become very
complex and in general, under full $\bm{k}_\perp$ integrations, wash out the corresponding
contributions. Consequently, the dominant terms at numerator and denominator are the Sivers
effect and the unpolarized term, which have the simplest azimuthal phases.

In order to perform the calculations, we strictly follow Refs.~\cite{Anselmino:2005ea,
Anselmino:2007fs} where the azimuthal asymmetries measured by the HERMES and COMPASS
Collaborations for semi-inclusive pion and charged hadron production have been used to
extract the Sivers function and, via a combined analysis of Belle data on
$e^+e^-\to\pi\pi+X$, the transversity distribution and the Collins fragmentation function. It
is important to recall that in SIDIS the Sivers and Collins contributions to the asymmetry
can be disentangled by considering suitable moments of the azimuthal distributions of the
observed hadrons.

In what follows, we will adopt the same Sivers and transversity
distributions and the Collins FF as in
Refs.~\cite{Anselmino:2005ea,Anselmino:2007fs}. Our aim is to
calculate the SSA for pion and kaon production in polarized $pp$
collisions, for the RHIC kinematical range. Notice that no fit is
performed,  we will simply compare our predictions to data on
unpolarized cross sections and SSAs from the STAR and BRAHMS
Collaborations. All TMD functions have a simple ansatz form in which
the $\bm{k}_\perp$ and lightcone fraction dependences are
factorized, with a flavor independent Gaussian shape for the
$\bm{k}_\perp$ term. For unpolarized PDFs and FFs the Gaussian
widths are fixed to $\langle k_\perp^2\rangle=0.25$ (GeV/$c)^2$ and
$\langle p_\perp^2\rangle=0.2$ (GeV/$c)^2$,
respectively~\cite{Anselmino:2005nn}. As in
Ref.~\cite{Anselmino:2005ea}, we adopt the MRST01 PDF
set~\cite{Martin:2002dr} and the Kretzer FF
set~\cite{Kretzer:2000yf}. Following Ref.~\cite{D'Alesio:2004up} we
fix the factorization scale to $\mu=\hat{p}_{cT}/2$, where
$\hat{p}_{cT}$ is the transverse momentum of the fragmenting parton
$c$ in the partonic c.m.~frame. Notice that, although our
calculations are performed at lowest order in pQCD, no additional
$K$-factors are introduced.

\begin{center}
\begin{figure*}
\epsfig{figure=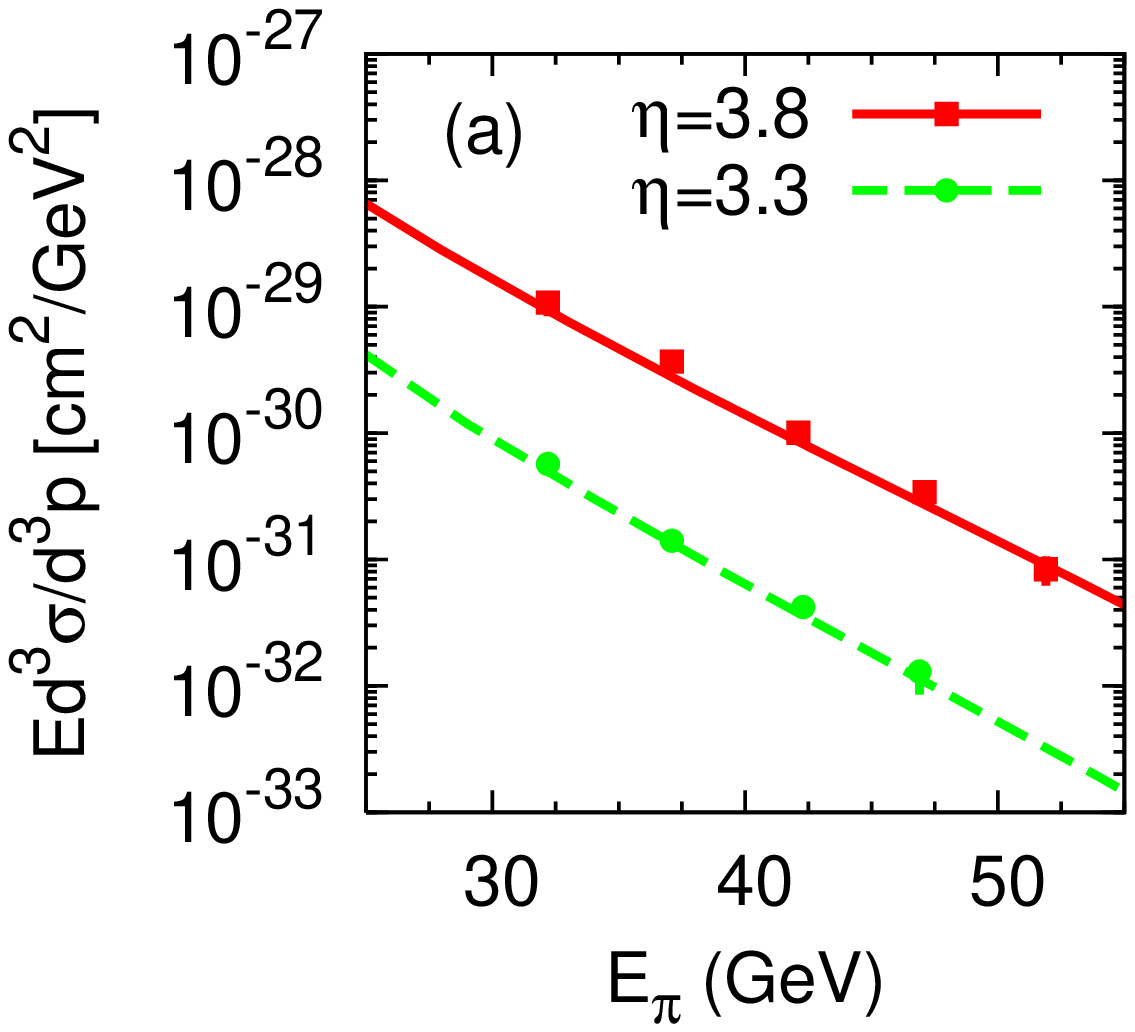,scale=0.44,angle=-90}
\epsfig{figure=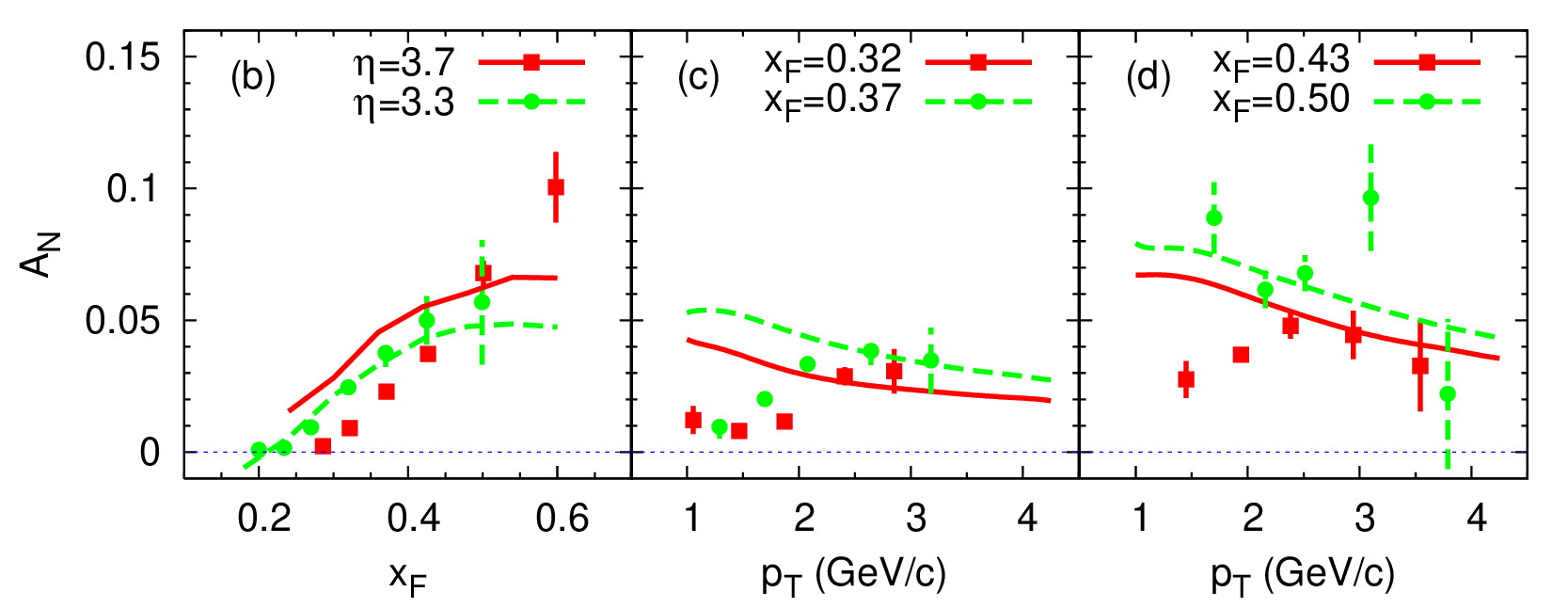,scale=0.42,angle=-90}
\caption{\label{star} Panel (a): Invariant differential cross
section for $pp \to \pi^0 +X$ at $\sqrt{s}= 200$ GeV and two
pseudorapidity values, $\eta=3.3$ and $\eta=3.8$, vs.~$E_\pi$. Data
are from STAR~\cite{Adams:2006uz}. Curves are obtained adopting the
unpolarized $k_\perp$-dependent PDFs and FFs of
Ref.~\cite{Anselmino:2005nn}. Panels (b), (c) and (d): $A_N$ for $pp
\to \pi^0 +X$ at $\sqrt{s}= 200$ GeV and two pseudorapidity values,
$\eta=3.3$ and $\eta=3.7$, vs.~$x_F$ (b), and at different $x_F$
values vs.~$p_T$, (c) and (d). Data are from
STAR~\cite{Nogach:2006gm}. Curves are obtained using the Sivers
functions as determined in Ref.~\cite{Anselmino:2005ea} by fitting
SIDIS data.} \vspace*{20pt}
\end{figure*}
\end{center}

In Fig.~\ref{star}, panel (a), we show the unpolarized cross section
for neutral pion production in $pp$ collisions, at $\sqrt{s}=200$
GeV and two different average pseudorapidities,
$\langle\eta\rangle=3.3$ and $\langle\eta\rangle=3.8$, as a function
of $E_\pi$. The corresponding SSAs, $A_N(p^\uparrow p\to\pi^0+X)$,
are plotted in panels (b), (c) and (d) as a function of $x_F$ at
fixed rapidity values, (b), and as a function of  $p_T$ at different
bins in $x_F$, (c) and (d). The curves are our predictions, compared
to STAR experimental data~\cite{Adams:2006uz,Nogach:2006gm}. In this
case the Collins effect is totally negligible.

In Fig.~\ref{brahms-pi}, (a) and (b) panels, we compare our
predictions to BRAHMS data on unpolarized cross sections for charged
pion production~\cite{Arsene:2007jd}, as a function of $p_T$ at
$\sqrt{s}=200$ GeV and two different rapidities, $y=2.95$ and
$y=3.3$. On the (c) and (d) panels, we plot the analogous results
and data for $A_N$~\cite{Lee:2007zzh}, as a function of $x_F$ and
for two different c.m.~scattering angles, $\theta=2.3^\circ$ and
$\theta=4.0^\circ$. In particular, the total contribution obtained
by adding the Sivers and the Collins effect (thick lines) is shown
together with the Collins contribution alone (thin lines).

\begin{center}
\begin{figure*}[h!]
  \hspace*{-50pt}
\epsfig{figure=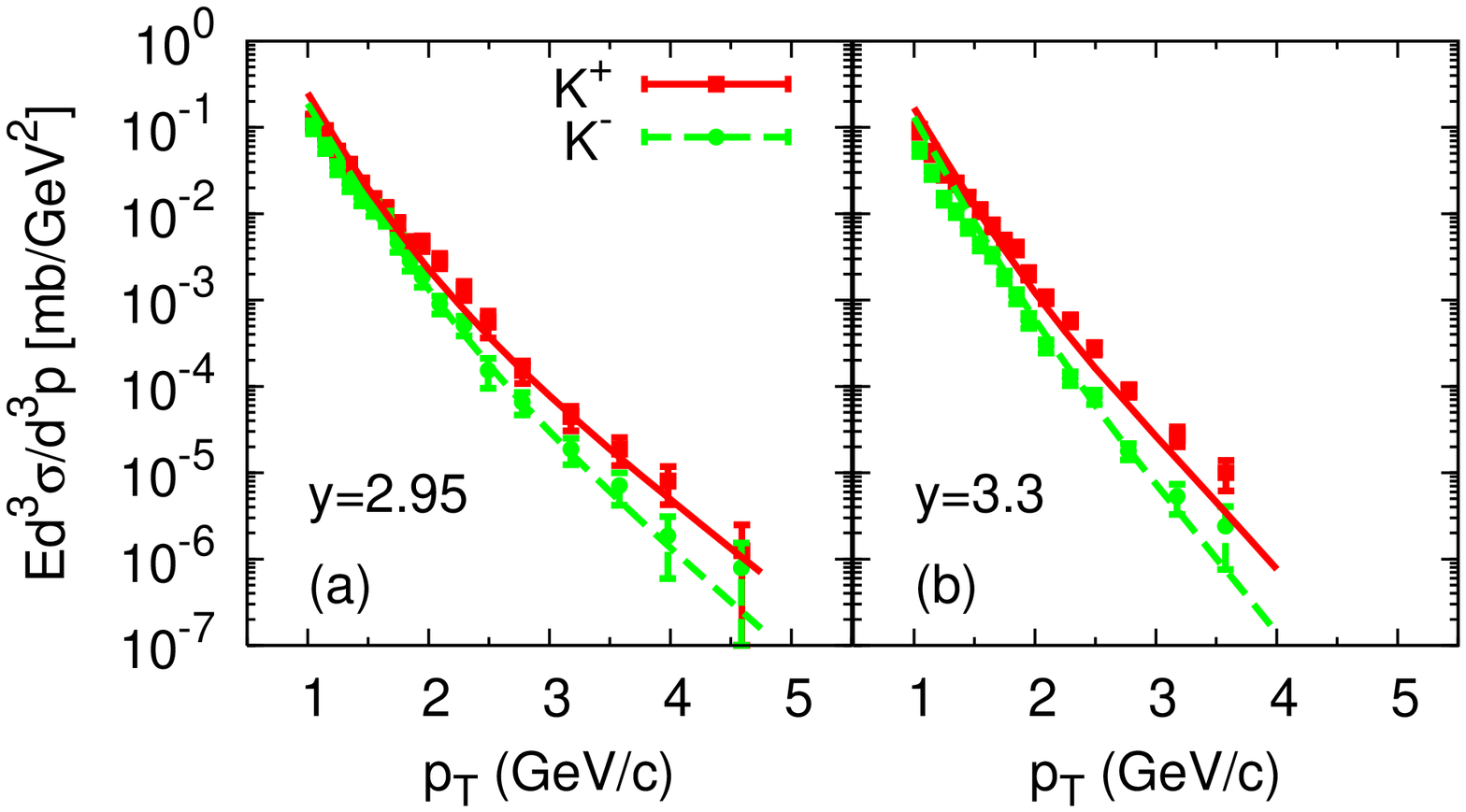,scale=0.42,angle=-90,bb=
  550 50 560 640}  \hspace*{15pt}
\epsfig{figure=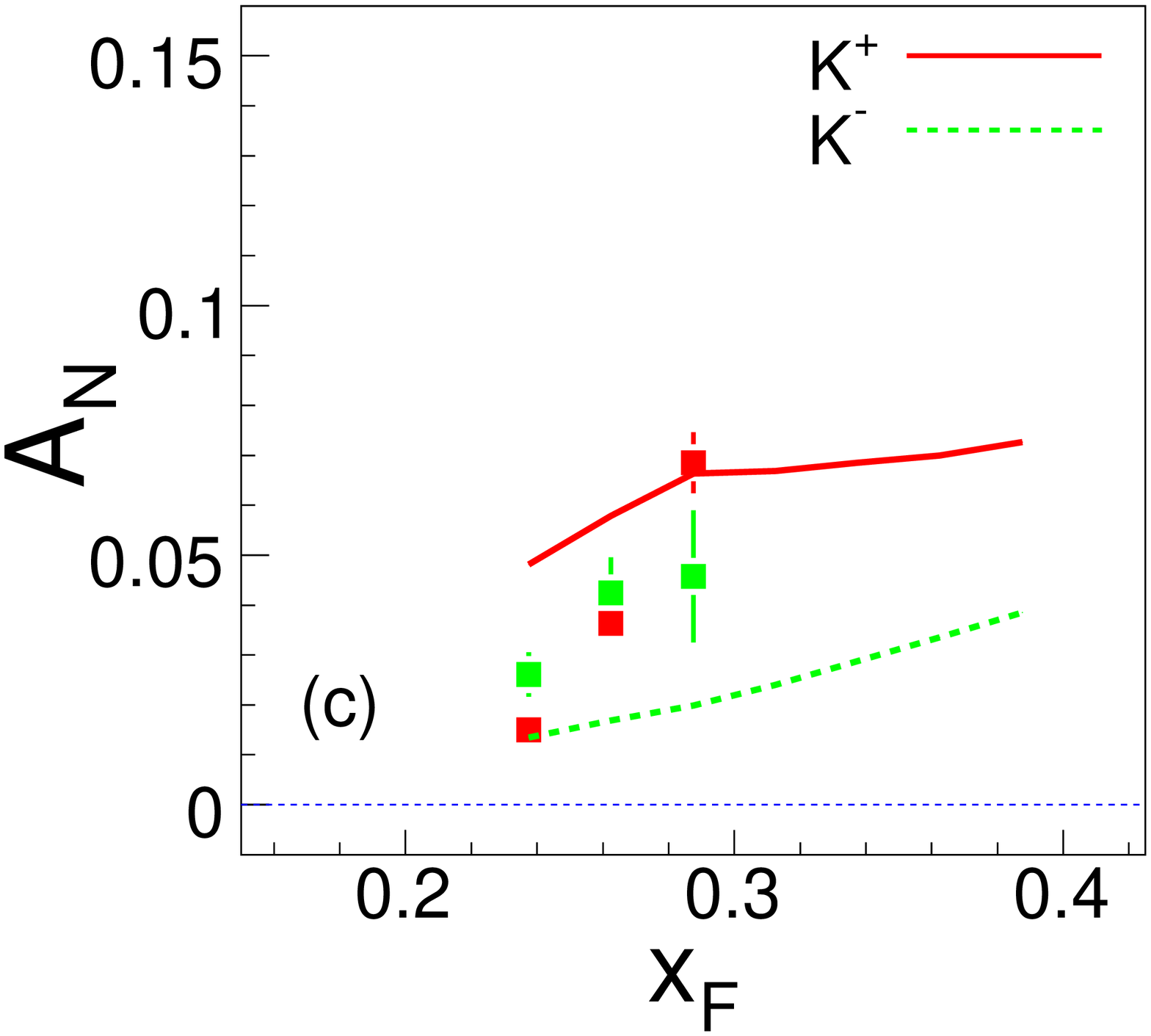,scale=0.24}
\caption{\label{brahms-K} Panels (a) and (b): Invariant differential
cross section  for $pp \to K^\pm +X$ at $\sqrt{s}= 200$ GeV and two
pseudorapidity values, $y=2.95$ and $y=3.3$, vs.~$p_T$. Data are
from BRAHMS~\cite{Arsene:2007jd}. Curves are obtained adopting the
same choices as in Fig.~1(a). Panel (c): $A_N$  for $pp \to K^\pm
+X$ at $\sqrt{s}= 200$ GeV and fixed scattering angle,
$\theta=2.3^\circ$, vs.~$x_F$. Data are from
BRAHMS~\cite{Lee:2007zzh}. Curves are obtained with the Sivers
function as determined in Ref.~\cite{Anselmino:2005ea} by fitting
SIDIS data.}
\end{figure*}
\end{center}

Finally, in Fig.~\ref{brahms-K} we compare our predictions for
unpolarized cross sections, (a) and (b), and SSAs, (c), to the BRAHMS
data for charged kaon production~\cite{Arsene:2007jd, Lee:2007zzh}.
Notice that for $A_N$ we consider only the Sivers effect as the
Collins contribution is small and the unknown kaon Collins functions
are still under study.

Let us comment on these results:\\
\nd 1) On the whole, with the exception of low $p_T$ STAR data, our predictions are in
remarkably good agreement, both in size and sign, with RHIC data. In particular, there are no
evident contradictions that could be interpreted as signals of sizable factorization and
universality breaking between SIDIS and $pp$ collisions. We believe this information may be
very useful in the context of the recent theoretical developments: in fact, although
factorization-breaking terms could in principle be there, it is not easy to infer how sizable
and significant their contributions should be for the processes and kinematical
configurations where data are presently
available. \\
 \nd 2) Although remarkable, one should not overestimate
the significance of our results. First of all, the HERMES and
COMPASS data on SIDIS used in the fits of
Refs.~\cite{Anselmino:2005ea, Anselmino:2007fs} are collected mainly
in the region of low-intermediate values of the Bjorken variable
($x_B \leq 0.3$). Therefore, the fits strictly constrain our
parameterization for the Sivers function only in the low $x$ region.
Consequently, with these parameterizations we are unable to
reproduce the low energy results on the SSA for charged pions at
large $x_F$ of the E704 collaboration. Similarly, with the average
$k_\perp$'s for PDFs and FFs extracted from SIDIS, we underestimate
the corresponding unpolarized cross sections by a factor which
cannot simply be attributed to pQCD $K$-factors. Further advances in
this direction would require a simultaneous fit of SIDIS and $pp$
data, which would give tighter constraints on the large
$x$ behaviour of our parameterizations.\\
\nd 3) In the color gauge invariant TMD approach of
Ref.~\cite{Bomhof:2006dp} gauge links result in numerical prefactors
which can be absorbed in generalized partonic cross sections. These
factors may be different for different processes, therefore breaking
factorization. No calculations have been performed within this
scheme for the single inclusive particle production in hadronic
collisions, but only for the double inclusive case. However, it is
possible that in the kinematical configurations considered here, the
dominant partonic contributions are those with positive prefactors,
close to one. Therefore, our findings cannot be interpreted as
contradicting the color gauge invariant approach; rather, they could
indicate that in these situations the generalized parton model and
the color gauge invariant approach give very similar results.
Unambiguously different predictions from the two approaches could
instead be found in SSAs in DY process and in photon-jet production
in $pp$ collisions, see e.g.~Refs.~\cite{Efremov:2004tp, Anselmino:2005ea, Bacchetta:2007sz}.\\
 \nd 4) For the $pp$ case (and for SIDIS at large transverse momentum
of the observed hadron, comparable with $Q$) there is a well known
alternative approach, the collinear twist-three
formalism~\cite{Qiu:1998ia}. Ref.~\cite{Kouvaris:2006zy} showed that
results comparable or better than those found in a TMD approach can
be obtained for SSAs in $pp$ collisions. Let us point out, however,
that in Ref.~\cite{Kouvaris:2006zy} all available $pp$ data were
included in the fit, while our results can be considered as genuine
predictions. Concerning the low energy E704
data, this approach faces similar problems both for SSAs and unpolarized cross sections.\\
5) The fit of SIDIS data used here was performed before data on
Sivers and Collins asymmetries for neutral pions and kaons were made
available by the HERMES collaboration. Also COMPASS results on
separated pions and kaons became available afterwards. This could
partially explain the poorer agreement with kaon production data
from BRAHMS. Nevertheless, let us stress that kaon data are much
more challenging in SIDIS itself, since they are crucially dependent
on the strange and $u$-, $d$-sea quark contributions to the Sivers
function. Notice that non-leading contributions in the fragmentation
functions are presently known with less accuracy also in the
collinear unpolarized case. Some progress in this direction has been
recently achieved and a new set of fragmentation functions is
available for kaons~\cite{deFlorian:2007aj} which differ
substantially from those adopted in this paper. A detailed study of
these effects is beyond the aim of this paper and requires a new
fitting procedure and an updated analysis, which will be discussed
elsewhere~\cite{Anselmino:2008xx}.

In conclusion, this phenomenological analysis shows that most of the
available high energy data on unpolarized cross sections and SSAs
for single inclusive particle production in SIDIS and in $pp$
collisions can be well reproduced in the framework of a generalized
parton model, with inclusion of spin effects and leading twist TMD
distributions, in particular the Sivers function.
 \onecolumngrid
%
%%%%%%%%%%%%%%%%%%%%%%%%%%%%%%%%%%%%%%%%%%%%%%%%%%%%%%%%%%%%%%%%%%%%%%%%%%%%%%
%%%%%%%%%%%%%%%%%%%%%%%%%%%%%%%%%%%%%%%%%%%%%%%%%%%%%%%%%%%%%%%%%%%%%%%%%%%%%%

%\vspace*{10pt}

%%%%%%%%%%%%%%%%%%%%%%%%%%%%%%%%%%%%%%%%%%%%%%%%%%%%%%%%%%%%%%%%%%%%%%%%%%%%%%
%%%%%%%%%%%%%%%%%%%%%%%%%%%%%%%%%%%%%%%%%%%%%%%%%%%%%%%%%%%%%%%%%%%%%%%%%%%%%%
\twocolumngrid

We remark that while SIDIS data on pion and charged hadron
production have been used for the fits, it is not so for the $pp$
case, where all our results are genuine predictions. Their agreement
with data seems to indicate that the role of possible
factorization-breaking terms may be marginal for the processes and
kinematical ranges considered here. We believe that this
phenomenological information may presently be very useful, given the
rapid development of this field of research. Our results might
indicate that our approach and the pQCD color gauge-link formalism
cannot be distinguished on the basis of these processes. One should
rather look at Sivers single spin asymmetries in DY processes, in
prompt photon or, as suggested in Ref.~\cite{Bacchetta:2007sz},
photon-jet production in $pp$ collisions. These reactions might also
be useful in disentangling our approach from the collinear
twist-three formalism.

In future, data at large $x_F$($x_B$) in $pp$ collisions(SIDIS)
would be very helpful in further constraining our parameterizations,
testing our model more severely. Moreover, new kaon data challenge
the phenomenology of SSAs, as they definitely involve the role of
sea quark TMD distributions. This issue will be studied in
Ref.~\cite{Anselmino:2008xx}.

\acknowledgments

We thank M. Anselmino and A. Prokudin for useful discussions and
J.~H.~Lee for his help in drawing Figs.~\ref{brahms-pi} and
\ref{brahms-K}. We acknowledge support of the European Community -
Research Infrastructure Activity under the FP6 ``Structuring the
European Research Area'' program (HadronPhysics, contract number
RII3-CT-2004-506078). M.B.~acknowledge partial support by MIUR under
Cofinanziamento PRIN 2006.

\bibliography{spiresppnp}

\end{document}